\documentclass[12pt,thmsa]{article}
\usepackage{amssymb}
\usepackage{sw20aip}

\input{tcilatex}
\begin{document}

\author{Todor Boyadjiev\thanks{%
Faculty of Mathematics and Computer Science, University of Sofia, 5, James
Bourchier Blvd., 1163 Sofia, Bulgaria. E-mail: \texttt{%
todorlb@fmi.uni-sofia.bg}} \and Nora Alexeeva\thanks{%
Department of Applied Mathematics, University of Cape Town, Rondebosh, 7700,
Wester Cape, South Africa. E-mail: \texttt{nora@maths.uct.ac.za}}}
\title{Josephson Lattices of the Optimal Size}
\date{\today}
\maketitle

\begin{abstract}
The stability of the bound states of the magnetic flux in a Josephson
resistive lattices is investigated numerically. It is shown that for a
simple relationship between the geometrical parameters of the lattice the
range of bias current is the widest.
\end{abstract}


\section{Formulation of the problem}

In the present paper we study numerically stationary spatially periodic
states of the magnetic flux in one-dimensional Josephson junctions, whose
dielectric layer represents a lattice of resistive inhomogeneities of finite
size \cite{1,2}.

The mathematical model of such junctions is described by the perturbed
stationary sin-Gordon equation 
\begin{equation}
-\phi _{xx}+j_{D}(x)\sin \phi +\gamma =0,\quad x\in (-R,R),  \label{1}
\end{equation}
supplemented with the periodic boundary conditions 
\begin{equation}
\phi (-R)=\phi (R),\quad \phi _{x}(-R)=\phi _{x}(R).  \label{2}
\end{equation}
Here $j_{D}(x)$ is given continuous periodic function which models the
distribution of the amplitude of the Josephson current along the junction; $%
2R$ is a given length of a spatial wave; $\gamma $ is the bias current.

Let us consider in more details the choice of the function $j_{D}(x)$. For
the homogeneous junctions we have $j_{D}(x)\equiv 1$. In an inhomogeneous
case the function $j_{D}(x)$ is defined by the number of the inhomogeneities
and their shape. The $\delta $-function model of inhomogeneities is the most
frequently used \cite{3,4,5,6,7,8,9}. In this paper we use inhomogeneities
of the shape of equilateral trapezium (see fig.~1). Such trapezium is
characterized by the width of its bottom $\mu $ and its top $\sigma =m\mu $.
In all cases the height of the trapezium equals to a dimensionless unit \cite
{3}.

For $m=1$ the inhomogeneity is a rectangular$\ $with base $\mu $. The more
realistic model of an inhomogeneity is one with $m>1$. The decay of the
amplitude of the Josephson current from $0$ to $1$ in such an inhomogeneity
takes place in the interval of length $\delta =\dfrac{m-1}{2}$ $\mu $ for
the particular choice of the width $\mu $ of the inhomogeneity.

Most of the calculations in this work were carried out for $m=1.5$, i.e. for 
$\delta =1/4$. We consider an ''infinite'' lattice consisting of the
described inhomogeneities with separation $\Delta $ between their center
positions. The simple relationship $\mu \leq \frac{\Delta }{m}$ between the
values of $\Delta $, $\mu $ and $m$ ensures that two adjacent
inhomogeneities do not overlap.

The lattice is symmetric when $\Delta -\sigma =\mu $; this yields $\mu =%
\dfrac{\Delta }{1+m}.$

To examine the stability of a certain solution $\phi (x)$ of nonlinear
boundary value problem (BVP) (\ref{1}), (\ref{2}) we solve the
Sturm-Liouville problem (SLP) 
\begin{equation}
-\psi _{xx}+q(x)\psi =\lambda \psi ,  \label{3}
\end{equation}
\begin{equation}
\psi _{x}(\pm R)=0,  \label{4}
\end{equation}
\begin{equation}
\dint \limits_{-R}^{R}\psi ^{2}(x)\,dx=1,  \label{5}
\end{equation}
where the potential 
\begin{equation}
q(x)=j_{D}(x)\cos \phi (x)  \label{6}
\end{equation}
is generated by this particular solution. The solution $\phi (x)$ is stable
if the minimal eigenvalue $\lambda _{\min }$ of the SLP with the
corresponding potential $q(x)$ given by (\ref{6}), is positive \cite{3}.

To solve numerically the nonlinear problems (\ref{1}), \ref{2} and \ref{3}
-- (\ref{5}) we use continuous analog of the Newton's method \cite{10} with
an optimal step \cite{11}. At each iteration we discretize the corresponding
linearized BVP using the spline-collocation second order scheme over a
nonuniform grid with condensed spacing in the vicinity of the
inhomogeneities. The detailed description of the algorithm and the numerical
scheme is given in \cite{1}.

\section{Discussion of the numerical results}

Each solution $\phi (x,\Delta ,\mu ,N_{I},\gamma )$ of BVP (\ref{1}), (\ref
{2}) depends on the geometrical parameters $\Delta $, $\mu $ and $N_{I}$,
and on the physical parameter $\gamma $ as well. Here $N_{I}$ is the number
of inhomogeneities, enveloped by a single wave of length $2R=N_{I}\Delta $.
Such waves exist in a real finite size junctions having $N_{I}$
inhomogeneities symmetric with respect to its center. Here $2R$ is the
length of the junction. We note that in this case the eq.~(\ref{1}) is
supplemented by the boundary conditions of Neuman type 
\begin{equation}
\phi _{x}(\pm R)=h_{B}.  \label{7}
\end{equation}
Here $h_{B}$ is the magnetic field at the boundaries of the junction.

The potential $q(x)$ of the SLP (\ref{3}) - (\ref{5}) is explicitly related
to the periodic solution of the BVP (\ref{1}), (\ref{2}) by eq.~(\ref{6}),
and, consequently, the corresponding eigenvalues and eigenfunctions depend
on the parameters $\Delta $, $\mu $ and $N_{I}.$ Thus the stability of the
bound states of the magnetic flux in the junction also depends on these
parameters.

Note that when the bias current $\gamma =0$ the eqs.~(\ref{1}), (\ref{2})
have ''vacuum'' (Meissner's) solutions of the form 
\[
\phi (x)=k\pi ,\quad k=0,\pm 1,\pm 2,\dots , 
\]
for every $N_{I}$. The solutions corresponding to even $k$ are stable while
the solutions corresponding to odd $k$ are unstable.

Let us consider the influence of the geometric parameters on the possible
bound states of the magnetic flux in the lattice.

For a junction with $N_{I}=1$, $2R=\Delta <2\pi $ and $\gamma =0$ we find
just two periodic solutions --- stable and unstable Meissner's solutions.
Such junction is ''short'' for periodic solutions, however, we should note
that there are nontrivial stable aperiodic solutions corresponding to the
eq.~(\ref{1}). The derivative $\phi _{x}(x)$ of one of these solutions is
marked by ''$\diamond $'' in fig.~2. For large enough values of the distance 
$\Delta $, nontrivial unstable periodic solutions exist in the junction. The
most interesting among them is the one, whose derivative is marked by ''$%
\Delta $'' in fig.~2. We will conditionally call this solution soliton. It
remains unstable in a very wide range of the values $\Delta $ and $\mu $;
this is demonstrated in fig.~3 and fig.~4. The second unstable solution,
marked by ''$\square $'' in fig.~2, exists in a real junction with one
inhomogeneity only for very large values of $h_{B}$ at the boundaries.

There is a periodic stable solution in the lattice in the case $N_{I}=2$.
Further we refer to this solution as ''main periodic soliton''. This
solution is marked by ''o'' in fig.~2 and fig.~5. If the two symmetric
inhomogeneities of a junction are separated by large enough distance $\delta 
$, then the periodic magnetic flux $\phi (x)$ is a result of the nonlinear
interaction between two aperiodic stable pinned at the inhomogeneities
structures - the fluxon and the antifluxon. Similarly, for the case $N_{I}=1$
the unstable solutions are the result of ''interactions'' between two
unstable fluxons, alienated from the inhomogeneity. For these reasons one
can make the conjecture that for the case $N_{I}=1$ there aren't any
nontrivial periodic stable solutions \cite{12}.

The solution, marked by ''$\nabla $'' in fig.~5 (the ''secondary'' periodic
soliton) is an example of weakly stable bound state, whose energy is
approximately four times greater then the energy of the main periodic
soliton. The stable soliton with small $\lambda _{\min }$ exists only for
very large values of the parameters $\Delta $ and $\mu $ of the lattice.
Both solitons have symmetric reflections about the axis $x$. Further we
refer to them as main periodic antisoliton and secondary periodic
antisoliton. The remaining two solutions in fig.~5, marked by ''$\square $''
and ''$\diamond $'' are unstable.

Fig.~6 shows the minimal eigenvalue $\lambda _{\min }$, corresponding to two
stable periodic solutions, as a function of the parameter $\mu $ when $%
N_{I}=2$ and bias current $\gamma =0$. Different curves correspond to
different values of the distance $\Delta $ between the inhomogeneities. It
is very important to note, that all the curves have maximum in a certain
point $\mu _{c}$, dependent on the parameter $\Delta $. This leads that in a
lattice with inhomogeneities separated by the distance $\Delta $ with width $%
\mu _{c}$, the time-dependent perturbations of the magnetic flux will decay
smoothly \cite{1}. The location of the points of maximum $\mu _{c}$ for all
the solutions depends on the value of $\Delta $ and for the main soliton (as
for main antisoliton) is equal approximately to $1.3$.

Apart from the periodic stable solutions in a real junction with two
inhomogeneities there are stable aperiodic bound states of the magnetic flux
when the boundary magnetic field $h_{B}=0$; this case corresponds to the
boundary conditions (\ref{7}).

The distribution of the aperiodic magnetic field $\phi _{x}(x)$ along the
junction is shown in fig.~7 (the symmetric reflections are not shown). Here
the values of the parameters are $R=6$, $N_{I}=2$, $h_{B}=0$, $\gamma =0$.
The marked by ''o'' solution is periodic, others are not. The energy of the
''complex'' bound state (periodic soliton), pinned at two inhomogeneities,
is much greater than the energy of the aperiodic solitons. The minimal
eigenvalue of the periodic soliton is smaller than the minimal eigenvalues
corresponding to the aperiodic bound states.

The minimal eigenvalue $\lambda_{\min}$ as a function of the parameter $\mu$
has an maximum for the aperiodic bound states as well as for the aperiodic
ones. This is demonstrated in fig.~8.

For comparison, the same curve (marked by ''$\diamond $''), corresponding to
the stable fluxon in the junction having one inhomogeneity, is shown in
fig.~9.

The curves $\lambda _{\min }(\mu )$, corresponding to the unstable periodic
solutions, are shown in fig.~9. Here the parameters of the lattice are $%
N_{I} $, $\mu =2.4$ and the bias current $\gamma =0.$ The stable main
periodic soliton transforms into a unstable solution at the point of
bifurcation $B_{0}$ (the unstable soliton is marked by ''$\square $'' in
fig.~5). The secondary periodic soliton is unstable in the lattice with so
''narrow'' inhomogeneities. The secondary soliton transforms into another
unstable periodic soliton at the point of bifurcation $B_{1}$ when the
second eigenvalue of the eqs.~(\ref{3}) - (\ref{5}) is also equal to zero.
The latter unstable soliton is marked by ''$\diamond $'' in fig.~5.
Increasing of the width $\mu $ of the inhomogeneities ''lifts'' the curve $%
\lambda _{\min }(\mu )$ upwards and, if $\mu $ is greater than a certain
critical value, the secondary periodic soliton becomes stable.

Let us consider the dependence of the bound states of the magnetic flux on
the bias current $\gamma $. It is well-known that in the inhomogeneous
junction the stable solutions lose their stability when the absolute value
of the bias current is increased. In the periodic case the process of
''destruction'' of the magnetic field by the bias current is demonstrated in
fig.~10. It can be seen\cite{13}, that the positive values of the current $%
\gamma $ ''compress'' the field in the center between two inhomogeneities,
while the negative values of $\gamma $ push it to the adjacent
inhomogeneities or, in the case of a real junction with two inhomogeneities,
to the boundaries of the junction.

The curves $\lambda_{\min}(\gamma)$ for all found stable solutions as well
as unstable, in the case $N_I=2$, $\Delta=6$, $\mu=2.4$ are demonstrated in
fig.~11. The curves, corresponding to the stable solutions, have their
symmetric reflections about the line $\gamma=0$. As before, $B_0$ denotes
the bifurcation points, corresponding to the roots of the equation $%
\lambda_{\min}(\gamma)=0$.

Let $\gamma _{a}$ and $\gamma _{b}$ are the roots of this equation (the
points $B_{0a}$ and $B_{0b}$ correspond to this roots in fig.~11 for the
main soliton/antisoliton). Let us designate the difference 
\[
\Delta \gamma =\gamma _{b}-\gamma _{a} 
\]
interval of stability of the main periodic soliton with respect to the bias
current $\gamma $. Generally, the value of $\Delta \gamma $ is a function of
the parameters $\Delta $ and $\mu $. The interval of stability $\Delta
\gamma $ for the main soliton as a function of the parameter $\mu $, when
the value of the distance $\Delta $ is fixed, is shown in the fig.~12. These
curves can be approximated by the cubic polynomials with coefficients,
depending only on $\Delta $, in the whole domain where $\mu $ is defined
physically. For a given $\Delta $ the points of maximum $\mu _{s}$ determine
the width of the inhomogeneities, so, that the soliton has the maximal
interval of stability $\Delta \gamma $ with respect to the current $\gamma $.

Note, that the shape of the curve $\mu_s(\Delta)$ can be fitted by a line $%
\mu_s(\Delta)=a\Delta+b$ with a good accuracy. This is the line, marked by
''o'' in fig.~13. The approximate values of the parameters of this line,
calculated numerically, are: $a\approx 0.254$, $b\approx -8.5.10^{-3}$.

Having in mind the above results we can construct an ''optimal'' junction
with geometric parameters related by 
\[
\mu _{c}=\mu _{s}(\Delta _{opt}). 
\]

In this case we have $N_{I}=2$, $\Delta _{opt}\approx 4\mu _{c}.$ Thus, the
''optimal'' distance between the inhomogeneities is $\Delta _{opt}\approx
5.2 $.

\section{Conclusions}

We show numerically that in the Josephson junctions with resistive
inhomogeneities the stable periodic bound states of the magnetic flux,
pinned on the inhomogeneities, have a certain minimal wavelength. The
dependence of the stable bound states on the width $\mu $ of the
inhomogeneities and the distance $\Delta $ between them is studied. In
particular, it is demonstrated that the curve $\lambda _{\min }(\mu )$ for a
fixed $\Delta $ has maximum $\mu _{c},$ i.e. in such a junction the
time-dependent perturbations of the magnetic flux decays smoothly. The
dependence of the interval of stability $\Delta \gamma $ on the parameter $%
\mu $ is also studied. For a certain relationship between the with of the
inhomogeneities and their separation (''optimal'' lattices) the interval of
stability $\Delta \gamma $ is the widest.

\section{Acknowledgments}

We thank Prof. S.N. Dimova (University of Sofia), Prof. I.V. Puzynin (JINR,
Dubna) and Prof. I.V. Barashenkov (University of Cape Town) for their useful
remarks.

This research was supported by the Bulgarian Ministry of Education, Science
and Technologies under the grant MM-425/94.

The work of NVA was partially supported by the FRD of South Africa.

\pagebreak

\section{Figure captions}

1. Geometrical model of a resistive inhomogeneity.

2. Some distributions of the magnetic field $\phi _{x}(x)$ along the
junction. The solution, marked by ''$\diamond $'', is stable aperiodic
soliton of the magnetic field in the case $N_{I}=1$; the solutions, marked
by ''$\square $'' and ''$\Delta $'' are unstable periodic (case $N_{I}=1$);
marked by ''o'' solution is stable periodic (case $N_{I}=2$).

3. Minimal Eigenvalues $\lambda _{\min }$ of the SLP (\ref{3}) - (\ref{5})
versus the distance $\Delta $ between the inhomogeneities for the solutions,
marked by ''$\square $'' and ''$\Delta $'' in fig.~2.

4. Minimal Eigenvalues $\lambda _{\min }$ of the SLP (\ref{3}) - (\ref{5})
versus the width $\mu $ of the inhomogeneities for the solutions, marked by
''$\square $'' and ''$\Delta $'' in fig.~2.

5. Stable (marked by ''$\diamond$'' and ''$\square$'') and unstable (marked
by ''$\diamond$'' and ''$\square$'') distributions of the magnetic field $%
\phi_x(x)$ along the junction with parameters $N_I=2$, $\Delta=9.5$, $%
\mu=2.4 $. Here $\gamma=0$.

6. Minimal eigenvalue $\lambda _{\min }$ of the SLP (\ref{3}) - (\ref{5})
versus the width $\mu $ of the inhomogeneities for the stable solutions from
fig.~5. Here the parameter $\Delta $ is fixed and the point $\mu _{c}$
corresponds to the ''optimal'' width.

7. Stable bound states of the magnetic field $\phi_x(x)$ corresponding to
the aperiodic boundary conditions (\ref{7}). Here $N_I=2$, $\Delta=6$, $\mu
=1$, $\gamma=0$, $h_B=0.$ The quantity $\Delta\phi=\phi(R)-\phi(-R)$ is the
full magnetic flux through the junction.

8. Minimal eigenvalue $\lambda _{\min }$ of the SLP (\ref{3}) - (\ref{5})
versus the width $\mu $ of the inhomogeneities for the stable solutions from
fig.~5. Here the parameter $\Delta $ is fixed.

9. Minimal eigenvalue $\lambda _{\min }$ of the SLP (\ref{3}) - (\ref{5})
versus the distance $\Delta $ between the inhomogeneities. $B_{0}$ and $%
B_{1} $ are the points of bifurcation.

10. Variation of the shape of the stable soliton as the bias current is
increased. Here $N_I=2$, $\Delta =6$, $\mu =2.4$.

11. Minimal eigenvalue $\lambda _{\min }$ of the SLP (\ref{3}) - (\ref{5})
versus the bias current $\gamma $. $B_{0a}$ and $B_{0b}$ --- for the
periodic soliton from fig.~7.

12. Interval of stability $\Delta\gamma=\gamma_b-\gamma_a$ versus the width $%
\mu$ of the inhomogeneities. The parameter $\Delta$ is fixed.

13. The width $\mu _{s},$ corresponding to the largest interval of stability
versus the distance $\Delta $ in the cases $N_{I}=2$ and $N_{I}=3.$

\pagebreak

\end{document}